\def\noi{\noindent}

\def\beq{\begin{equation}}
\def\eeq{\end{equation}}
\def\bmulteq{\begin{eqnarray}}
\def\emulteq{\end{eqnarray}}

\def\m{m}
\def\M{M}

\def \Im{{\rm Im}}
\def \Re{{\rm Re}}

\def \pperpl{$P_l^\bot$}
\def \pperpmu{$P_\mu^\bot$}
\def \pperptau{$P_\tau^\bot$}
\def \pperpavg{$\overline{P_\mu^\bot}$}
\def \pperpavgl{$\overline{P_l^\bot}$}
\def \pperpavgtau{$\overline{P_\tau^\bot}$}

\def\pperp{$P^{\perp}$}

\def\ep{\varepsilon}

\def \epsmu{\epsilon_{\mu\alpha\beta\gamma}}

\def\naive{na\"\i ve}

\def\abs #1{\left|#1\right|}

\def\plet #1 {\left\lgroup \matrix{#1} \right\rgroup}

\documentstyle[12pt]{article}

\begin{document}

\begin{flushright}
TRI-PP-94-11\\
hep-ph/9403389\\
Mar. 1994
\end{flushright}

\renewcommand{\baselinestretch}{1.5} \small\normalsize

\

\centerline{\Large $CP$ violating polarizations}
\centerline{\Large in semileptonic heavy meson decays}

\centerline{Robert Garisto}

\centerline{\it
TRIUMF,
4004 Wesbrook Mall,
Vancouver, B.C.,
V6T 2A3,
Canada
}

\

\

We study the $T$-violating lepton transverse polarization (\pperpl)
in three body semileptonic heavy meson decays to pseudoscalar mesons and to
vector mesons.  We calculate these polarizations
in the heavy quark effective limit,
which simplifies the expressions considerably.
After examining constraints from $CP$ conserving (including $b \rightarrow
s \gamma$) and $CP$ violating processes, we find that in $B$ decays,
\pperp\ of the muon in multi-Higgs doublet models
can be of order $10\%$,
while \pperp\ of the $\tau$ can even approach unity.
In contrast, \pperpmu\ in $D$ decays is at most 1.5\%.
We discuss possibilities for detection of \pperpl\ at current and future
$B$ factories.
We also show that \pperpl\ in decays to vector mesons, unlike in decays
to pseudoscalars, can get contributions from left-right models.
Unfortunately, \pperpl\ in that case is proportional to $W_L$-$W_R$ mixing,
and is thus small.


\newpage
\section {Introduction}

The Standard Model (SM) has thus far met with incredible experimental success.
Nevertheless, many hypothetical extensions to the SM
remain phenomenologically viable.
Since new physics  often provides new sources of $CP$ violation (CPV),
one good way to search for such extensions
is to consider $CP$ violating observables which are negligible
in the SM, but which can have large contributions from
other sources of CPV.

A major barrier to any candidate for such an observable is the upper bound
on the electric dipole
moment of the neutron, $d_n$,
which is now around $10^{-25}$ e cm \cite{dn expt}.
The SM explanation for CPV, the
Cabbibo-Kobyashi-Maskawa (CKM)  mechanism \cite{KM},
has come to be accepted by many
as the source of $CP$ violation in the neutral $K$ sector
not only because it predicts $\epsilon$ to be in the right range,
but also because it predicts $d_n$ to be negligible \cite{Barr Marciano}.
As the upper bound on $d_n$ has plummeted, many potential explanations for
$\epsilon$ from other sources have run aground, and thus it is more difficult
to find observables which have good prospects of detecting CPV beyond the
SM.

One such observable is the
transverse polarization of the lepton in
semileptonic $K_{\mu3}$ decays \cite{Kane Frere}, \pperpl,
which is the $T$-violating projection of
the lepton spin onto the normal of the decay plane, $i.e.$
$P_l^\perp \sim \vec s_l \cdot (\vec k \times \vec p)$
\cite{Lee Wu}, where $\vec k$ and
$\vec p$ are decay product momenta.
It arises from the interference between two amplitudes with non-zero
relative phase.
In practice, one measures the asymmetry between the number of particles
parallel and anti-parallel to the normal of the decay plane,

\beq
P^\perp_l = { N_\uparrow - N_\downarrow \over
 N_\uparrow + N_\downarrow}.
\label{N over N}
\eeq

\noi
There are several advantages to using such a semileptonic
$CP$ violating observable.  First, semileptonic decays occur through a
single SM diagram at tree level,
so \pperpl\ is negligible in the SM
\cite{Zhit}.
Thus a non-zero \pperpl\ is a signal for new physics.
Second, theoretical uncertainties in semileptonic decays are much smaller
than in purely hadronic decays.
Finally, \pperpl\ in semileptonic decays comes from both
the quark and lepton sectors, so that
purely hadronic or purely leptonic $CP$ violating observables, such as $d_n$
or $d_e$, do not necessarily strongly constrain \pperpl\
\cite{foot semileptonic}.
In fact, there exist reasonable models for which \pperpl\ in $K_{\mu3}$ decays
can be of order $10^{-2}$--$10^{-3}$, consistent with all other constraints
\cite{Kmu3,Belanger}.
Such values are well within reach of experiments.
The last measurements of \pperpl\ were done
at Brookhaven National Lab on
$K^+ \rightarrow \pi^0 \mu^+ \nu_\mu$ decays.  Their combined
result was \cite{Blatt etal}

\beq
P^\perp_\mu \left( K^+ \rightarrow \pi^0 \mu^+ \nu_\mu \right)
= -1.85 \pm 3.60 \times 10^{-3},
\label{pperpK expt}
\eeq

\noi
which implies a $95\%$ confidence upper bound of about $0.7\%$.
There is also
an experiment currently under construction at KEK \cite{KEK proposal}
which hopes to push this bound down by a factor of ten \cite{Kuno}.

In this paper we consider \pperpl\ in heavy meson decays of the type
$M\rightarrow m\,l\,\nu_l$ and $M\rightarrow m^*\,l\,\nu_l$,
where $M$ and $m$ are pseudoscalar mesons, and $m^*$ is a vector meson.
\pperpl\ has been studied in decays to pseudoscalars \cite{Cheng82,Soni},
but not in decays to vector mesons.
We derive expressions for \pperpl\ in $M\rightarrow m^*\,l\,\nu_l$ decays,
as well as in $M\rightarrow m\,l\,\nu_l$ decays,
in the heavy quark effective limit.
This greatly simplifies our results.
One can even obtain analytic expressions for \pperpavgl,
the polarization averaged over all kinematical variables.

In decays to pseudoscalars ($M\rightarrow m\,l\,\nu_l$),
\pperpl\ is sensitive only to spin 0 effective Lagrangians
\cite{Kane Frere,foot tensor},
which makes it a good tool for probing non-SM Higgs physics
\cite{Kmu3}.  We find that this holds for \pperpl\ in
decays to {\it longitudinally polarized} vector mesons
($M\rightarrow m_L^*\,l\,\nu_l$), but that \pperpl\ in
decays to {\it transversely polarized} vector mesons
($M\rightarrow m_{T1}^*\,l\,\nu_l$ and $M\rightarrow m_{T2}^*\,l\,\nu_l$)
is sensitive only to new $V$ and $A$ physics, such as left-right models.
Unfortunately, \pperpl\ in that case is proportional to $W_L$-$W_R$ mixing,
which is constrained to be small.
However, in the former case,
multi-Higgs doublet models yield encouraging results, even after
imposing $CP$ conserving and $CP$ violating constraints.  There are
reasonable models in which \pperptau\ in
$B \rightarrow D \tau \nu$ decays can even approach unity.

Section 2 lists the form factors needed for our calculation.
We consider contributions to \pperpl\
from multi-Higgs doublet models in Section 3, and from
left-right models in Section 4.
Possibilities for detecting \pperpl\  in various decay modes are discussed
in Section 5.

\section {Form Factors}

{}From Lorentz invariance and basic symmetry considerations, we
can write the hadron matrix elements (HME)
for decays of pseudoscalar mesons ($M$)
to pseudoscalar ($m$)  and vector mesons ($m^*$) as

\bmulteq
&\left<\m(k)|\, V_\mu \, | \M(K)\right> =&
f_+\, (K+k)_\mu + f_-\,(K-k)_\mu\, , \nonumber \\
&\left<\m(k)|\, A_\mu \, | \M(K)\right> =& 0 , \nonumber\\
&\left<\m^*(k,\ep^*_\lambda)|\, V_\mu \, | \M(K)\right> =&
{i V_1 \over M} \left( \epsmu\, \ep_\lambda^{*\,\alpha}K^\beta k^\gamma \right)
 , \nonumber\\
&\left<\m^*(k,\ep^*_\lambda)|\, A_\mu \, | \M(K)\right> =&
A_1 M {\ep^*_\lambda}_\mu + {A_2 \over M}(\ep^*_\lambda \cdot K) (K+k)_\mu
\nonumber\\
&&+  {A_3 \over M}(\ep^*_\lambda \cdot K) (K- k)_\mu
 ,
\label{VA mom FF}
\emulteq

\noi
where for $M^+$ decay, $V_\mu = \bar D \gamma_\mu U$ and
 $A_\mu = \bar D \gamma_\mu \gamma^5 U$ ($U$ and $D$ are the appropriate
up- and down-type quarks for $M$ and $m$).
The axial vector HME for $M \rightarrow m$ is zero because there
is no way to form an axial vector with just $K^\alpha$ and $k^\beta$.
We have used $M$ and $m$ to represent both a meson and its mass.
The form factors $f_\pm$, $V_1$, and $A_{\rm 1-3}$
are functions of $(K\cdot k)$ and $r\equiv m/M$.
Here $\lambda$ is the polarization index. We will refer to
the $m^*$ longitudinal
polarization by the label $\lambda = L$, and the two transverse polarizations
by the label $\lambda = T_1,\ T_2$, for $\vec\ep_\lambda$ in the decay plane
and
perpendicular to the decay plane, respectively.

{}From these vector and axial vector HME, one can derive scalar
and pseudoscalar HME using the
Dirac equation \cite{Cheng82}:

\bmulteq
&\left<\m(k)|\, S \, | \M(K)\right> =&
{- M^2 \over m_D - m_U} \left(f_+ (1 - r^2) + f_- t \right) , \nonumber\\
&\left<\m(k)|\, P \, | \M(K)\right> =& 0 , \nonumber\\
&\left<\m^*(k,\ep^*_\lambda)|\, S \, | \M(K)\right> =& 0,  \nonumber \\
&\left<\m^*(k,\ep^*_\lambda)|\, P \, | \M(K)\right> =&
{- M \over m_D + m_U}\, (\ep^*_\lambda \cdot K) (A_1 + A_2 (1-r^2) +A_3 t)
,
\label{SP}
\emulteq

\noi
where for $M^+$ decays, $S= \bar D\,U$, and $P= \bar D \gamma^5 U$,
and $t \equiv (K-k)^2/M^2$.
The masses $(m_D,m_U)$ are $(m_b,m_c)$ in $B$ decays and $(m_s,m_c)$
in $D$ decays.
The middle two parity-odd matrix elements in
(\ref{SP}) are zero because there
is no way to form a pseudoscalar using only $K^\alpha$, $k^\beta$ and
${\ep_\lambda}^\gamma$.  Note that the factor $(\ep^*_\lambda \cdot K)$
implies that
$\left<\m^*(k,\ep^*_\lambda)|\, P \, | \M(K)\right> $ is non-zero
only for longitudinally polarized vector mesons.

Recently there has been a lot of interest in {\it heavy quark effective
theory} (HQET), which considers the limit $M$, $m \rightarrow \infty$.
The principle tenet of HQET is that $v_\mu$ ($v'_\mu$),
the four-velocity of $M$ ($m^{(*)}$), is unchanged by QCD corrections
\cite{GeorgiHQET}.  Thus it makes sense to write the HMEs in terms
of velocity \cite{Luke}:

\bmulteq
&\left<\m(v')|\, V_\mu \, | \M(v)\right> =&
\sqrt{Mm} \left( \xi_+ (v + v')^\mu + \xi_- (v-v')^\mu \right)
, \nonumber \\
&\left<\m(v')|\, A_\mu \, | \M(v)\right> =& 0 , \nonumber\\
&\left<\m^*(v',\ep^*_\lambda)|\, V_\mu \, | \M(v)\right> =&
i \sqrt{Mm} \,
\xi_{V_1} \left( \epsmu\, \ep_\lambda^{*\,\alpha}{v'}^\beta v^\gamma
\right)
 , \nonumber\\
&\left<\m^*(v',\ep^*_\lambda)|\, A_\mu \, | \M(v)\right> =&
\sqrt{Mm} \Bigl( \xi_{A_1} (1 + v \cdot v') {\ep_\lambda^*}_\mu
- \xi_{A_2} (\ep_\lambda^* \cdot v) v_\mu \nonumber\\
&&- \xi_{A_3} (\ep_\lambda^* \cdot v) v'_\mu \Bigr),
\label{VA vel FF}
\emulteq

\noi
{}From (\ref{VA mom FF}) and
(\ref{VA vel FF}), one can derive relations between the form
factors \cite{Neubert92}:

\bmulteq
&f_\pm =& \pm {1 \over 2 \sqrt{r}} \left( (1 \pm r) \xi_+ -
( 1 \mp r) \xi_- \right) \ \rightarrow \pm {1 \pm r \over 2 \sqrt{r}} \xi,
   \nonumber\\
& V_1 =& - {1 \over \sqrt{r}} \xi_{V_1}\
 \rightarrow - {1 \over \sqrt{r}} \xi,
 \nonumber\\
& A_1 =&  {x + r \over \sqrt{r}} \xi_{A_1}\ \rightarrow
{x + r \over \sqrt{r}} \xi,
\nonumber\\
& A_2 =& - {1 \over 2 \sqrt{r}} \left( \xi_{A_3} + r \xi_{A_2} \right )
\ \rightarrow - {1 \over 2 \sqrt{r}} \xi,
\nonumber\\
& A_3 =& + {1 \over 2 \sqrt{r}} \left( \xi_{A_3} - r \xi_{A_2} \right )
 \rightarrow + {1 \over 2 \sqrt{r}} \xi ,
\label{FF relations}
\emulteq

\noi
where $x\equiv (K \cdot k) /M^2 = r\, v \cdot v'$.
In the $M$, $m \rightarrow \infty$ limit,
$\xi_+ = \xi_{V_1}=\xi_{A_1} = \xi_{A_3} = \xi$ and $\xi_-=\xi_{A_2}=0$,
so that all the form factors can be written in terms of
the Isgur and Wise function, $\xi(x)$ \cite{Isgur Wise}.
Note that the HMEs are normalized so $\xi(x)$ is equal to 1
at zero recoil ($x=r$ or $v \cdot v' =1$)
\cite{Falk etal}.
Specific forms for $\xi(x)$ are listed in the Appendix.

\section {Higgs models}

\subsection{Transverse Polarization}

As we said, semileptonic pseudoscalar decays to pseudoscalar mesons,
$M \rightarrow m \, l \, \nu$, and to longitudinally polarized vector
mesons, $M \rightarrow m^*_L \, l \, \nu$, can arise only from
the interference of new scalar physics with the SM.
In this section, we
consider contributions to $P^\perp(M \rightarrow m\, l\, \nu_l)$
and $P^\perp(M \rightarrow m^*_L\, l\, \nu_l)$ from
models with charged Higgs scalars. Other types of contributions
are possible, such as from scalar leptoquarks
\cite{Kane Frere,Kmu3,Belanger}.

T. D. Lee first proposed that $CP$ could be violated via phases in a
model with two Higgs doublets \cite{Lee}.  This idea was
refined by S. Weinberg with the elimination of
flavor changing neutral currents (FCNC) in a model with three Higgs doublets,
using a symmetry to ensure that
only one Higgs doublet couples to each right-handed fermion field;
what is commonly referred to as
natural flavor conservation (NFC) \cite{Weinberg NFC}.
There are various other ways to avoid the FCNC problem
\cite{Georgi Nanopoulos,Hall Weinberg},
but for simplicity, we concentrate on models where NFC is either exact, or
partially broken \cite{Liu Wolf}.
We will assume that the CKM phase is non-zero, so we do
$not$ impose strong constraints on CPV in the Higgs sector
from $\epsilon$.
Even if $CP$ is broken only spontaneously, a non-zero CKM phase can arise
after integrating out super-heavy fields, so we see no reason to take it zero.

We are interested in the interference of a charged Higgs boson with the
SM $W$ boson, so one need only parametrize an effective Lagrangian for the
charged Higgs coupling to fermions.  In a model with $N$ charged scalar
fields, one obtains a Lagrangian in terms of the $N-1$ physical charged
Higgs bosons \cite{Cheng82}:

\begin{displaymath}
 - {\cal L}_{H^+} = {1 \over v} \sum_{i=1}^{N-1} \Bigl[
\alpha_i \bar U_L V_L M_D D_R H_i^+  +  \beta_i \bar U_R M_U V_L D_L H_i^+
\nonumber
\end{displaymath}
\beq
+ \gamma_i \bar N_L M_E E_R H_i^+ \Bigr] + H.c.
\label{Higgs L}
\eeq

\noi
Here v is the SM Higgs VEV,
$v = (4 G_F / \sqrt{2})^{-1/2}\simeq 174$GeV;
$U$, $D$, $N$, and $E$ are fields for
the up quarks ($U^T = (u\ c\ t)$), down quarks, neutrinos and charged leptons;
$M_D$, $M_U$ and $M_E$ are the diagonal mass matrices; and
$V_L$ is the CKM matrix.

If the coefficients $\alpha_i$, $\beta_i$ and $\gamma_i$ are complex,
the interference between the charged Higgs and $W$ boson amplitudes in
Fig. 1 produces a $T$-violating transverse polarization of the lepton.
Since the $H^+$ amplitude is proportional to the matrix elements in
(\ref{SP}), one gets contributions to \pperpl$(M \rightarrow m^* l \nu)$
only for decays in which the $m^*$ is longitudinally polarized.
This means that if one can veto decays with transversely polarized
$m^*$'s, the denominator in (\ref{N over N}) will be reduced while
the numerator will remain unchanged, leading to a larger polarization.

Let us evaluate \pperpl\ in terms of the velocity dependent form factors.
Then we can take the heavy quark effective limit, which allows us to
write \pperpl\ with only one form factor, $\xi(x)$.
We calculate \pperpl\ for semileptonic pseudoscalar decays to
pseudoscalar mesons, to longitudinally polarized vector mesons,
and to unpolarized vector mesons in this limit:

\beq
P^\perp_l (x) \left( M \stackrel{H^+}{\rightarrow} m l \nu \right)=
C_{H^+} {3\pi\over 4}\,
{(1-r^2) (x+r) (x^2-r^2) \sqrt{t}  \over
(1+r)^2 x_1^3 }
\, {\xi(x)^2 \over \xi(x)^2},
\label{Px S}
\eeq
\beq
P^\perp_l (x) \left(M \stackrel{H^+}{\rightarrow} m_L^* l \nu \right) =
C_{H^+} {3\pi\over 4}\,
{ (1-r^2) (x+r) (x^2-r^2) \sqrt{t}  \over
(1-r)^2 (x+r)^2 x_1 }
\, {\xi(x)^2 \over \xi(x)^2},
\label{Px L}
\eeq

\beq
P^\perp_l (x) \left(M \stackrel{H^+}{\rightarrow} m^* l \nu \right) =
C_{H^+} {3\pi\over 4}\,
{(1-r^2) (x+r) (x^2 -r^2) \sqrt{t}  \over
(1-r)^2 (x+r)^2 x_1 + 4 t (x+r) x x_1 }
\, {\xi(x)^2 \over \xi(x)^2}.
\label{Px LT}
\eeq

\noi
We list the full expressions with general form factors in the Appendix.
Note that we have already integrated \pperpl\ over
one kinematical variable ($(K \cdot p)/M^2$) so that
\pperpl\ is only a function of
the remaining kinematical variable $x$
(where $x=(K \cdot k)/M^2$ and $x_1 \equiv \sqrt{x^2 - r^2}$).
This integration gives the factor $3 \pi/4$ in (\ref{Px S})--(\ref{Px LT}).
For $M^+ \rightarrow \bar m^{0\,(*)} l^+ \nu_l$ and
$M^0 \rightarrow  m^{-\,(*)} l^+ \nu_l$ decays,
the new physics coefficient is given by:

\beq
C_{H^+} = {M m_l \over M_W^2}\, \sum_i^{N-1}\, {M_W^2 \over M_{H_i}^2}
\left( {m_D \over m_D \mp m_U} \Im \alpha_i \gamma_i^* +
{m_U \over m_D \mp m_U} \Im \beta_i \gamma_i^* \right),
\label{CH}
\eeq

\noi
while $C_{H^+}$ for the $CP$ conjugate decays has the opposite sign
\cite{Okun}.
Here $m_l$, $m_U$ and $m_D$ are the lepton and current quark masses
specific to each decay,
and $M_{H_i}$ and the coefficients $\alpha_i$, $\beta_i$ and
$\gamma_i$ come from the effective Lagrangian (\ref{Higgs L}).
The $-$ ($+$) applies to $M \rightarrow m l \nu$
($M \rightarrow m^* l \nu$) decays. Since $m_U > m_D$ in $D$ decays,
it follows that \pperpl$(D^+ \rightarrow \bar K^0 l^+ \nu)$ has the
opposite sign as
$P_l^\perp(K^+ \rightarrow \pi^0 l^+ \nu)$,
$P_l^\perp(B^+ \rightarrow \bar D^{0(*)} l^+ \nu)$, and
$P_l^\perp(D^+ \rightarrow \bar K^{0*} l^+ \nu)$.
It also means that $C_{H^+}$ in the decays to $m^*$ are somewhat suppressed
over those to $m$ when $m_D$ and $m_U$ are of the same order, as
in $B$ decays.

We have neglected all lepton mass effects in the denominator of
(\ref{Px S})--(\ref{Px LT}).
For $l=\mu$, this is always a very small effect. In $l=\tau$ decays,
it changes our results only qualitatively when \pperptau$\sim 1$,
$i.e.$, when $H^+$ effects are important in the denominator of
(\ref{N over N}).  In that case, it might be possible to see new physics
effects in changes to the branching ratio of $B \rightarrow D^{(*)} \tau \nu$.

To estimate the size of \pperpl\ in various models, we must integrate over
the remaining kinematical variable $x$. In an experiment, one generally
measures the overall asymmetry in (\ref{N over N}), rather than measuring
\pperpl$(x)$ for each $x$ and then averaging.  So we must integrate the
numerator and denominator of (\ref{Px S})--(\ref{Px LT}) separately:

\bmulteq
\overline{P^\perp_l} \left( M \stackrel{H^+}{\rightarrow} m l \nu \right)=&&
C_{H^+} {3\pi\over 4}\, { I_\perp \over I_S}, \nonumber\\
\overline{P^\perp_l} \left( M \stackrel{H^+}{\rightarrow} m^*_L l \nu
\right)=&&
C_{H^+} {3\pi\over 4}\, { I_\perp \over I_L}, \nonumber\\
\overline{P^\perp_l} \left( M \stackrel{H^+}{\rightarrow} m^* l \nu \right)=&&
C_{H^+} {3\pi\over 4}\, { I_\perp \over I_L + I_T},
\label{P avg scalar}
\emulteq

\noi
where $I_\perp$, $I_S$, $I_L$ and $I_T$ are integrals
of the kinematics in (\ref{Px S})--(\ref{Px LT}).
Unfortunately, this means we must know something about the overall form factor
$\xi(x)$.  In the Appendix, we list two possible parametrizations
for $\xi(x)$: a relativistic oscillator model, and a monopole approximation.
$\overline{P^\perp_l}$ in decays to pseudoscalars in
these models differs by at most $15\%$ for $r$ in
the region of interest ($r>0.25$ for all the decays we study),
and considerably less for decays to vector mesons.
If we set the monopole parameter $\rho$  equal to 1 in (\ref{xi monopole}),
we can obtain analytic expressions
for $\overline{P^\perp_l}$ in terms of $r$.  We list the corresponding
$I$'s in the Appendix.
{}From Fig. 2, we see that
choosing $\rho=1$ instead of $1.2$ (in order to obtain analytic expressions)
changes $\overline{P^\perp_l}$ by only a few percent (for $r >0.25$).
Even na\"\i vely dividing out $\xi(x)^2$ from the numerator
and denominator of (\ref{Px S})--(\ref{Px LT}) gives results which
(for $r >0.25$)
differ by $30\%$, or much less, from the other parametrizations of $\xi(x)$.

\subsection{Constraints}

For the purposes of placing constraints on \pperpl, we make two simplifying
assumptions.
First, we take the $\alpha_i$,
$\beta_i$ and $\gamma_i$ to be flavor diagonal.  This strictly holds only
in models with NFC, so Higgs models without NFC may have somewhat weaker,
more model-dependent bounds
\cite{foot Klmumu}.
Second, we will assume that the lightest charged Higgs
mass eigenstate, $h^+$, gives the dominant contribution, so that we can
drop the subscript $i$ on the coefficients $\alpha$, $\beta$ and $\gamma$.
In 3HDMs, $\Im \alpha_1 \gamma_1^* = - \Im \alpha_2 \gamma_2^*$
and $\Im \beta_1 \gamma_1^* = - \Im \beta_2 \gamma_2^*$,
so in that case we are simply making the replacement
$M_{H_1^+}^{-2} - M_{H_2^+}^{-2} \rightarrow M_{h^+}^{-2}$. This has virtually
no effect on $CP$ violating constraints, because they have the same behavior,
and the $CP$ conserving constraints tend to require a large splitting
between $M_{H_1^+}$ and $M_{H_2^+}$ anyway.

We want to constrain $C_{H^+}$, which now depends upon
$\Im\alpha\gamma^*$, $\Im\beta\gamma^*$, $M_W^2/M_{h^+}^2$, and the
masses involved with $M$ and $m^{(*)}$.
In the general case (given our two
assumptions), we can bound $\Im\alpha\gamma^* \, M_w^2/M_{h^+}^2$ directly
from the experimental upper bound on
$P^\perp_\mu(K^+ \rightarrow \pi^0 \mu^+ \nu_\mu)$ of
$0.7\%$ \cite{Blatt etal} to obtain

\beq
\abs{\Im \alpha\gamma^*} \, {M_W^2 \over M_{h^+}^2} < 730.
\label{Im ag}
\eeq

\noi
Since $m_U$ is small, $P^\perp_\mu(K^+ \rightarrow \pi^0 \mu^+ \nu_\mu)$
is insensitive to $\Im\beta \gamma^*$.  The best we can do is
to use $\abs{\Im \beta\gamma^*} < \abs{\beta} \cdot \abs{\gamma}$.
{}From the bounds placed upon $\abs{\beta}$ and $\abs{\gamma}$ by
\cite{Grossman}, we obtain

\beq
\abs{\Im \beta\gamma^*} \, {M_W^2 \over M_{h^+}^2} < 160 {M_W \over M_{h^+}}
< 285,
\label{Im bg}
\eeq

\noi
{}From (\ref{Im bg}), one sees that the upper bound on
$\Im\beta\gamma^* \, M_W^2/M_{h^+}^2$ decreases with increasing $M_{h^+}$
and is at its maximum when $M_{h^+}$ is at the model independent lower bound
of $M_Z/2$.  We can use (\ref{Im ag}) and (\ref{Im bg}) in
(\ref{P avg scalar}) to obtain upper bounds on \pperpavgl\ for various decays.
Our results are summarized in the first column of Table 1.

Let us now specialize to the case of 3HDMs.  The $CP$ violating coefficients
can be written \cite{Kmu3}

\bmulteq
\Im \alpha \gamma^* &= {1 \over 2} \sin 2 \theta_3 \, \sin\delta
{v v_u \over v_d v_e}, \nonumber\\
\Im \beta \gamma^* &= {1 \over 2} \sin 2 \theta_3 \, \sin\delta
{v v_d \over v_u v_e}, \nonumber\\
\Im \alpha \beta^* &= {1 \over 2} \sin 2 \theta_3 \, \sin\delta
{v v_e \over v_u v_d},
\label{Im 3HDM}
\emulteq

\noi
where $v_u$, $v_d$, and $v_e$ give mass to the up quarks, down quarks and
charged leptons, respectively.  $\theta_3$ ($\delta$) is a free,
$CP$ conserving ($CP$ violating) parameter of the model. For convenience,
let us define

\beq
\kappa \equiv \abs{\sin 2 \theta_3 \, \sin\delta} \, {M_W^2 \over M_{h^+}^2},
\label{def kappa}
\eeq

\noi
so that $\Im\alpha\gamma^* \, M_W^2/M_{h^+}^2$ and
$\Im\beta\gamma^* \, M_W^2/M_{h^+}^2$ are just given in terms of
$\kappa$ and the VEVs.  The relations in (\ref{Im 3HDM})
are not enough by themselves
to better the constraints given by (\ref{Im ag}) and (\ref{Im bg}), so
we consider specific models.

A common assumption is that the three VEVs are all of the same order,
$i.e.$, $v_u \simeq v_d \simeq v_e$.  We refer to this as the
{\it VEV Equality} (VE) model.
In the VE model, all three $CP$ violating coefficients are of
order one, and \pperpavg\ will be quite small.
But with one VEV for each type
of massive fermion, this need not be the case.
Since fermion masses are proportional to the VEVs as well as the Yukawa
couplings, it is quite reasonable to suppose that the hierarchy in
the fermion masses lies in
the VEVs, and not the Yukawa couplings \cite{Kmu3}.
Suppose the third family Yukawa couplings are of the same order.
Then one has $v_u:v_d:v_e \sim  m_t:m_b:m_\tau$, which implies
that

\beq
\abs{\Im \alpha\gamma^*} \, {M_W^2 \over M_{h^+}^2} \sim
{m_t^2 \over m_b m_\tau} \kappa,
\label{Im ag kappa}
\eeq

\noi
so that \pperpavgl\ need not be small \cite{Kmu3}.
We will refer to this as the {\it VEV Hierarchy} (VH) model.
While the VH model provides a reasonable justification for considering
large ratios of VEVs, it does not solve all the mass hierarchy problems.
We view the VE and VH models as two reasonable extremes, much in the same
way  that the range $1$ to $m_t/m_b$ is considered for ``$\tan\beta$''
in 2HDMs.

For simplicity, we define the VEVs in the VE model to be identically
equal, and in the VH model to have the ratio $m_t:m_b:m_\tau$ exactly.
Since (\ref{def kappa}) implies $\kappa < 3.2$ or so,
$P^\perp_\mu(K^+ \rightarrow \pi^0 \mu^+ \nu_\mu)$ does not put
any further constraints on $\kappa$ in the VE model.
However, the VH model can reach the upper bound on
$P^\perp_\mu(K^+ \rightarrow \pi^0 \mu^+ \nu_\mu)$,
and one needs $\kappa < 0.5$.  We now must consider if there
are any other constraints on $\kappa$ which would force \pperpavgl\ to
be small.

As we said in the Introduction, the most stringent constraint on
CPV often comes from the electric dipole moment of the neutron, $d_n$.
The purely hadronic coefficient $\Im \alpha\beta^*$ is very constrained
by $d_n$ \cite{Kmu3},
and in the VE model we find that we need $\kappa<1.2$.
However, in the VH model, $\Im\alpha\gamma^* /\Im\alpha\beta^*$ is
large, and the upper bound on $\kappa$ is only about $6$, which is ten times
weaker than the $K_{\mu3}$ bound.  This is a consequence of
the semileptonic decay---only quark-lepton CPV is enhanced in the VH model.

$CP$ conserving processes may also constrain \pperpavgl.
Consider the inclusive decay $b \rightarrow s \gamma$,
whose branching ratio is now bounded below $5.4 \times  10^{-4}$
at the 95\% confidence level by the CLEO collaboration
\cite{CLEO}.
This FCNC decay occurs via a one loop diagram in the SM with a branching
ratio of $3$-$4\times 10^{-4}$ \cite{Misiak}.
In 2HDMs, the charged Higgs contribution adds constructively with the
SM contribution, and one can bound the charged Higgs mass to be above
about $450$ GeV \cite{Barger Hewett}.
One would like to generalize this result to 3HDMs.  The amplitude
for $b \rightarrow s \gamma$ (at the $W$ mass scale)
can be written \cite{Pokorski}

\bmulteq
A &=& F_1\left({m_t^2 \over M_W^2}\right) +
{1\over3} \abs{\beta_i}^2 F_1\left({m_t^2\over M_{H_i^+}^2}\right)
+ \Re\alpha_i\beta_i^* \, F_2\left({m_t^2\over M_{H_i^+}^2}\right)
\nonumber\\
&+& i \Im\alpha_i\beta_i^* \, F_2\left({m_t^2\over M_{H_i^+}^2}\right),
\label{BSG ampl}
\emulteq

\noi
where the sum over $i$ runs from 1 to $N-1$, and one can show that
$\beta_i \beta_i^* = (v^2 - v_u^2)/v_u^2$,
$\Re \alpha_i \beta_i^*  =1$,
and $\Im \alpha_i \beta_i^*  =0$.
In the SM, only the first term is non-zero.
For $N=2$, we recover the 2HDM limit, {\it i.e.}
 $(\abs{\beta}^2, \Re\alpha\beta^*, \Im \alpha\beta^*) \rightarrow
(v_d^2/v_u^2, 1, 0)$.
In 3HDMs, one can have cancellations between the pieces
as long as $H_1^+$ and $H_2^+$ are not degenerate in mass.
It turns out that for both the VE and VH models, $\Re\alpha_1 \beta_1^*$
can be less than zero, so that for sufficiently large $M_{H_2^+}$,
there is no bound from $B(b \rightarrow s \gamma)$ on $M_{H_1^+}$.
For $M_{h^+} \sim M_W$ (or smaller),
$\sin 2 \theta_3 \sin\delta$  must be somewhat smaller than one
\cite{foot SUSY BSG},
but this is not enough to better the constraints
on $\kappa$ we have derived thus far.

There are also constraints from
$B(b \rightarrow s \gamma)$ on $\Im \alpha\beta^*$
\cite{Pokorski,Nir Grossman}, which in turn
constrains \pperpl\ via (\ref{Im 3HDM}).
Since the last term in (\ref{BSG ampl})
is purely imaginary, it does not destructively interfere with the other
terms, so that the contribution from $\Im \alpha\beta^*$
to $B(b \rightarrow s \gamma)$ is always positive.
However, even for $M_{h^+} \sim M_Z/2$, one can only
bound $\Im\alpha\beta^* < 2$
\cite{Nir Grossman},
which is satisfied in both the VE and VH models.
Since the CLEO observation of
$B(B \rightarrow K^* \gamma)= (4.5 \pm 1.5 \pm 0.9 ) \times 10^{-5}$
\cite{CLEO}
effectively sets a $lower$ limit on
$B(b \rightarrow s \gamma)$ of about $10^{-4}$, the
constraint on $\Im \alpha\beta^*$ from $b \rightarrow s \gamma$
will never be able to strongly constrain \pperpl\ in these models.

Finally, we note that the VE and VH models
give specific predictions for $\Im \beta \gamma^*$ (see
(\ref{Im 3HDM})), and in both cases it must be less than 2.
In general 3HDMs, one cannot improve upon the bound in
(\ref{Im bg}), though large $\Im \beta \gamma^*$ would require
small $v_u/v_d$ as well as very large $v/v_e$, which is
not as appealing theoretically as
either the VE or VH models.

In Table 1, we summarize the maximum values for \pperpavgl\ allowed in
the VE (VH) model, with a bound of
$\kappa < 1.2$ ($\kappa<0.5$) coming from the upper bound on
$d_n$ (\pperpl\ in $K$ decays).


\section{Left-Right Models}

Decays to vector mesons, $M\rightarrow m^* l \nu$, have one more 4-vector
than $M\rightarrow m l \nu$ decays
with which to construct hadronic matrix elements.
The $m^*$ polarization vector lets
us construct both a vector and an axial vector current
(see (\ref{VA mom FF})),
allowing a non-zero $V$ and $A$ interference term.  The upshot
is that $P^\perp_l(M\rightarrow m^* l \nu)$
gets contributions from spin 1 effective $CP$ violating Lagrangians
as well as those of spin 0.

Let us therefore consider left-right models
\cite{Mohaptra review},
whose charged gauge boson couplings to fermions can
be parametrized by the following effective Lagrangian:

\begin{displaymath}
 - {\cal L}_{W^+} =
{g_L \over \sqrt{2}}
\left[
\bar U_L \gamma_\mu V_L D_L  +  \bar N_L \gamma_\mu E_L \right] W_L^{+\, \mu}
\nonumber
\end{displaymath}
\beq
+ {g_R \over \sqrt{2}}
\left[ \bar U_R \gamma_\mu V_R D_R \right] W_R^{+\, \mu}
 + H.c. ,
\label{LR L}
\eeq

\noi
where $V_R$ is the right-handed CKM matrix.
We neglect right-handed currents coupled to leptons because they yield
polarizations proportional to $m_\nu$.
This means that \pperpl\ must arise from the interference of the
SM $W_L$ diagram and a diagram containing $W_L$-$W_R$ mixing (see Fig. 1)
\cite{Kane Frere}.
We define the mixing angle $\zeta$ by

\beq
\plet{W_1 \cr W_2\cr} =
 \pmatrix{\cos \zeta & \sin \zeta \cr
-\sin \zeta & \cos \zeta \cr }
\plet{W_L \cr W_R\cr } ,
\label{xiLR}
\eeq

\noi
where $W_1$ and $W_2$ are the two mass eigenstates.
The interference between $V$ and $A$ HMEs
vanishes for longitudinally polarized $m^*$'s,
so \pperpl\ is only non-zero for transversely polarized $m^*$'s.
Further, the numerator of (\ref{N over N}) has the same magnitude,
but opposite sign, for $m^*$'s with $T1$ and $T2$ polarizations.
Therefore, the polarization in the sum of decays
to both transversely polarized $m^*$ states,
$P^\perp_l(M\rightarrow m^*_T l \nu)$, is identically zero, and we must
consider \pperpl\ for either $T1$ or $T2$.
We again write \pperpl\ in the heavy quark effective limit,

\bmulteq
P^\perp_l (x) \left(M \stackrel{W_R^+}{\rightarrow} m^*_{T1} l \nu \right)
&=&
C_{W_R^+} {3\pi\over 4}\,
{(x+r) (x^2 -r^2) \sqrt{t}  \over
2 t (x+r)  x_1 (x - r/2)}
\, {\xi(x)^2 \over \xi(x)^2} \nonumber\\
P^\perp_l (x) \left(M \stackrel{W_R^+}{\rightarrow} m^*_{T2} l \nu \right)
&=&
-C_{W_R^+} {3\pi\over 4}\,
{(x+r) (x^2 -r^2) \sqrt{t}  \over
2 t (x+r)  x_1 (x + r/2)}
\, {\xi(x)^2 \over \xi(x)^2}
\label{Px LR}
\emulteq

\noi
and list the full expressions in the Appendix.  The
coefficient,

\beq
C_{W_R^+} = 2 \, { m_l \over M}\, \tan \zeta \
\Im \left( {g_R {V_R}^{UD} \over g_L {V_L}^{UD}} \right),
\label{C WR}
\eeq

\noi
depends upon the $W_L$-$W_R$ mixing angle $\zeta$,
the left and right CKM elements $V_L^{ij}$ and $V_R^{ij}$ ($i,j = U,D$), and
gauge coupling constants $g_L$ and $g_R$.

We can find an averaged polarization by
integrating the  numerator and denominator of (\ref{Px LR}) over $x$:

\bmulteq
\overline{P^\perp_l}
\left( M \stackrel{W_R^+}{\rightarrow} m^*_{T1} l \nu \right)=&&
C_{W_R^+} {3\pi\over 4}\, { I_\perp/(1-r^2) \over I_{T1}}, \nonumber\\
\overline{P^\perp_l}
\left( M \stackrel{W_R^+}{\rightarrow} m^*_{T2} l \nu \right)=&&
- C_{W_R^+} {3\pi\over 4}\, { I_\perp/(1-r^2) \over I_{T2}}.
\label{P avg LR}
\emulteq

\noi
We again use the $\rho=1$ monopole expression for $\xi(x)$, which results
in the $I_{T1}$ and $I_{T2}$ listed in the Appendix.
We have normalized the $I$'s so that $I_{T1} + I_{T2} = I_T$.
Fig. 3 shows that
using $\rho=1$ (to obtain an analytic expression)
instead of $1.2$ is a good approximation since $\xi(x)^2$
appears in both the numerator and denominator in (\ref{Px LR}).

Let us consider constraints on \pperptau\ in $B_{\tau3}$ decays.
Our Lagrangian in (\ref{Higgs L}) gives a {\it tree level} contribution to
$\epsilon'$ \cite{Chang}, and we can relate \pperpl\ and
$\epsilon'$.  If $\Im(V_R^{UD}/V_L^{UD})$ is roughly the same order for
all $UD$, then $\overline{P^\perp_\tau} \sim 10^{-2} (\epsilon'/\epsilon)$,
which is tiny.  It is  in principle possible that
$\Im(V_R^{ud}/V_L^{ud}) \simeq \Im(V_R^{us}/V_L^{us})\simeq 0$
while $\Im(V_R^{cb}/V_L^{cb})\sim 1$, which gives
\pperpavgtau\ $\sim 2 \zeta$.
Nevertheless, $\abs{\zeta}$ is constrained to be less than about 6\%
from $\mu$ decays \cite{mu zeta}, and less than about 2\%
from $b \rightarrow s \gamma$
\cite{BSG zeta},
so that we can bound \pperpavgtau\ to be less than about 4\%.


\section{Discussion}

Let us consider the various decay modes.
In particular, we discuss
whether one should  study charged or neutral decays, of $B$ or $D$ mesons, to
pseudoscalar or vector mesons, with $l=\mu$ or $l=\tau$.

Technically, the transverse polarization, \pperpl,
 is {\it motion reversal violating},
which is equivalent to $T$ violation only in the absence of final
state interaction (FSI) effects
\cite{Sachs}.
This is irrelevant in charged decays,
$e.g.$ $M^+ \rightarrow \bar m^0 l^+ \nu_l$,
because they have only one charged decay product, and FSIs are negligible.
In neutral decays, $e.g.$ $M^0 \rightarrow m^- l^+ \nu_l$, there are two
charged particles in the final state, so one can expect FSI effects of
order $\alpha_{EM}/\pi$ \cite{Brondine}.
For this reason, measurements of \pperpl\ in $K_{\mu3}$ decays are done
on the $K^+ \rightarrow \pi^0 \mu^+ \nu_\mu$ mode.  But if the experimental
sensitivity to \pperpl\ in a given decay is only at the percent level,
one can study decays of neutral mesons as well.
Actually, since both $B$ and $D$ mesons are produced in pairs,
one must be able to determine the charge of the lepton
(because \pperpl\ flips sign for the $CP$ conjugate decay) so that one
effectively measures the asymmetry

\beq
A_{CPV} \equiv {1 \over 2} \left[P^\perp_l(M \rightarrow \bar m l^+ \nu_l) -
P^\perp_l(\bar M \rightarrow m l^- \bar \nu_l)\right],
\label{A CPV}
\eeq

\noi
which is a true $CP$ violating observable.
Since FSI effects cancel in $A_{CPV}$, charged decays are in principle not
preferable to neutral decays.

{}From Table 1, it is clear that $B$ decays give larger \pperpavgl\ than
$D$ decays.
One can see from (\ref{CH}) that this has two causes: $M_D$ is
smaller than $M_B$, and the heavier quark mass in $D$ decays, $m_c$,
is proportional to $\Im \beta\gamma^*$ instead of $\Im \alpha\gamma^*$.
The former coefficient is more constrained than the latter,
and models in which  $\Im \beta\gamma^*$ is large tend to be
less theoretically appealing. For example, in 3HDMs, one would need
$v_u/v_d$ to be small while $v/v_e$ is very large.

Let us estimate the number of decays necessary to see a $5\sigma$ signal of
\pperpavgl\ with the maximum allowed values in the general case (column 1 of
Table 1).  We use $N = 25 k / \overline{P^\perp_l}^2$, and take $k\sim 10$.
One needs about $2.5\times 10^4$ $B\rightarrow D \mu \nu$ decays, which
(\naive ly)
translates into $2 \times 10^6$ $B$'s (including $B^\pm$, $B^0$, and
$\bar B^0$). In $B\rightarrow D^* \mu \nu$ decays, \pperpavgl\ is
6 times smaller (if one does $not$ veto decays to transversely polarized
$m^*$'s), but the branching ratio is about 3 times larger, so one needs
about 12 times as many $B$'s as in $B\rightarrow D \mu \nu$ decays to see
a signal.
By contrast,
one needs about $1.3 \times 10^6$ ($2.8 \times 10^7$) $D \rightarrow K\mu \nu$
decays to observe \pperpavgl\ for the maximum value in the
general case (VH model), which \naive ly requires
$4 \times 10^7$ ($10^9$) $D$'s.

To observe \pperpl\ of a muon, one needs to stop the muon so it can decay.
At a $symmetric$ $B$ factory, such as CESR or DORIS II,
the muon in $B\rightarrow D \mu \nu$ will have momentum of up to
2.3 GeV, which would require perhaps $1.3 kg/cm^2$ of material ($e.g.$
$\sim 4.5m$ of Al) to stop it
\cite{Kuno Bryman}. Unfortunately, even $2.5\times 10^4$
$B\rightarrow D \mu \nu$ decays is probably out of reach of either machine.
Stopping muons would be more difficult at the asymmetric SLAC $B$ factory,
since the muon momenta will be higher in the lab frame, but if
it could be accomplished, the luminosity
should be sufficient to see a $10\%$ polarization.
One could consider measuring \pperpmu\ at a hadron collider,
where the number of $B_{\mu3}$ and $D_{\mu3}$ decays would be much greater,
but the hurdle of stopping the muon would need to be overcome.

A better possibility may be $B_{\tau3}$ decays, because one can have
$\overline{P^\perp_\tau}\sim 1$.
One needs perhaps 250 $B\rightarrow D \tau \nu$ decays,
and 3000 $B\rightarrow D^* \tau \nu$ decays to see a $5 \sigma$ signal.
Both of these are probably out of reach of the existing symmetric machines,
but should be no problem for the SLAC $B$ factory.  Unlike muons,
taus do not need to be stopped, and one can measure the polarization of
the $\tau$ from its decay spectrum \cite{Bernabeu}.
In $\tau^\pm\rightarrow \pi^\pm  \nu_\tau$ decays, for example, the
decay width has the behavior $d\Gamma \sim 1 \mp \vec P_{\tau^\pm}
\cdot \hat p_\pi  \sim 1 -   P^\perp_{\tau^+} \cos \theta$
\cite{Tsai},
where $\vec P_{\tau^\pm}$ is the polarization vector of the $\tau^\pm$,
$\hat p_\pi$ is a unit
vector in the pion direction, and $\theta$ is the angle of $\hat p_\pi$
from the normal of the $B$ decay plane.
The main problem with $B_{\tau3}$ decays at the SLAC $B$ factory
may lie in defining the decay plane, since the $B$'s do not decay at rest,
in which case we may have underestimated $k$
\cite{Blatt etal,KEK proposal}.

Finally, we note that \pperpl\ from left-right models is probably unobservable
at the SLAC $B$ factory.
In addition to the small values for \pperpavgl\ required by the bounds
on $W_L$-$W_R$ mixing, one needs to measure the polarization of $m^*$ as well
as of $l$, so that our $k$ is perhaps $100$ or more.
For \pperpavgtau\ $\sim 4\%$, one needs more than $10^6$
$B\rightarrow D^* \tau \nu$ decays.

We have derived expressions for the transverse polarization of the lepton
in semileptonic meson decays, in the heavy quark effective limit.
Reasonable multi-Higgs models can
give a muon polarization in $B$ decays of order $10\%$ and a tau
polarization of order unity.  Both of these should be within the luminosity
reach of the SLAC $B$ factory, though the tau polarization has the advantage
of not requiring a stopper.
Should a non-zero signal be observed, implying the existence of
physics beyond the Standard Model, the best place to study \pperpl\
would be at a high luminosity {\it symmetric} $B$ factory.

\centerline{\bf\Large Acknowledgments}

I sincerely appreciate helpful conversations with
G. Kane, J. Bernab\'eu, J. N. Ng, D. Bryman and Y. Kuno.
This work was supported in part by a grant
from the Natural Science and Engineering Research Council of Canada.


\newpage

\centerline{\bf Table 1}

\begin{center}
\begin{tabular}{||c||c|c|c||}\hline
       & $\abs{\,\overline{P^\perp_l}\,}$ & $\abs{\,\overline{P^\perp_l}\,}$ &
$\abs{\,\overline{P^\perp_l}\,}$ \\
       &      general     &        VE        &   VH        \\
Decay  &       case      &        model      &  model      \\
\hline\hline
$K^+ \rightarrow \pi^0\, \mu^+\, \nu_\mu$  &
 $0.7\%$ & $1 \times 10^{-5}$ & $0.7\%$ \\
\hline
$D^+ \rightarrow \bar K^0\, \mu^+\, \nu_\mu$  &
 $1.4\%$ & $4 \times 10 ^{-5}$ & $0.3\%$\\
\hline
$D^+ \rightarrow \bar K^{0*}_L\, \mu^+\, \nu_\mu$  &
 $0.51\%$ & $2 \times 10^{-5}$ & $0.1\%$ \\
\hline
$D^+ \rightarrow \bar K^{0*}\, \mu^+\, \nu_\mu$  &
  $0.27\%$ & $1 \times 10^{-5}$ & $0.06\%$\\
\hline
$B^+ \rightarrow \bar D^0\, \mu^+\, \nu_\mu$  &
10\% & $2 \times 10 ^{-4}$ & $9.7\%$ \\
\hline
$B^+ \rightarrow \bar D^{0*}_L\, \mu^+\, \nu_\mu$ &
 $3.3\%$ & $6\times 10^{-5}$ & $2.9\%$ \\
\hline
$B^+ \rightarrow \bar D^{0*}\, \mu^+\, \nu_\mu$ &
$1.7\%$ & $3\times 10 ^{-5}$ & $1.5\%$ \\
\hline
$B^+ \rightarrow \bar D^0\, \tau^+\, \nu_\tau$  &
 $\sim 1$ & $3\times 10^{-3}$ & $\sim 1$ \\
\hline
$B^+ \rightarrow \bar D^{0*}_L\, \tau^+\, \nu_\tau$ &
 $\sim 55\%$ & $1 \times 10^{-3}$ & $\sim 50\%$ \\
\hline
$B^+ \rightarrow \bar D^{0*}\, \tau^+\, \nu_\tau$ &
 $\sim 29\%$ & $0.5 \times 10^{-3}$ & $\sim 25\%$ \\
\hline
\end{tabular}
\end{center}

\medskip

\begin{quote}
{\bf Table 1:}
Maximum values of the transverse polarization ($\overline{P^\perp_l}$)
for various decay modes
due to SM interference with
charged Higgs bosons in the general case and in two specific models.
$K^{0*}_L$ ($D^{0*}_L$) refers to longitudinally polarized $K^{0*}$'s
($D^{0*}$'s).
For the VE (VH) model, we use $\kappa < 1.2$ ($\kappa < 0.5$) as derived in
the text. Numbers of order one are approximate since we neglect $H^+$
effects in the denominator of (\ref{N over N}).
\end{quote}


\newpage
\appendix
\centerline{\bf\Large Appendix}

So far, we have used the heavy quark effective limit, in
which $\xi_+(x) = \xi_{V_1}(x)=\xi_{A_1}(x) = \xi_{A_3}(x) = \xi(x)$, and
$\xi_-(x)=\xi_{A_2}(x)=0$.
For completeness, we list the expressions for the polarization without that
simplification:

$$
P^\perp_l (x) \left( M \stackrel{H^+}{\rightarrow} m l \nu \right)=
C_{H^+} {3\pi\over 4}\,
{x_1^2 \sqrt{t}  \over x_1^3}\, \times
$$
$$
{\left[(1+r) \xi_+(x) + (1-r) \xi_-(x)\right]
  \over
\left[(1+r) \xi_+(x) + (1-r) \xi_-(x)\right]^2}\, \times
$$
\beq
\left[(1-r) (x+r) \xi_+(x) + (1+r) (x-r) \xi_-(x)\right] ,
\label{Px S full}
\eeq

$$ 
P^\perp_l (x) \left(M \stackrel{H^+}{\rightarrow} m_L^* l \nu \right) =
C_{H^+} {3\pi\over 4}\,
{x_1^2 \sqrt{t} \over (x+r)^2 x_1 } \times
$$ 
$$ 
{\left[\xi_{A_1}(x) (x+r) (x - r^2) - (\xi_{A_3}(x) + r \xi_{A_2}(x)) x_1^2
\right]
\over
\left[\xi_{A_1}(x) (x - r^2) - (\xi_{A_3}(x) + r \xi_{A_2}(x)) (x-r)\right]^2}
\times
$$ 
\beq
\left[ \xi_{A_1}(x) (x+r) - {1\over2}(\xi_{A_3}(x) + r \xi_{A_2}(x))
(1-r^2) + {1\over2} (\xi_{A_3}(x) - r \xi_{A_2}(x)) t \right]
\label{Px L full}
\eeq

$$ 
P^\perp_l (x) \left(M \stackrel{H^+}{\rightarrow} m^* l \nu \right) =
C_{H^+} {3\pi\over 4}\,
{x_1^2 \sqrt{t} \over (x+r) x_1 } \times
$$ 
$$ 
\left[\xi_{A_1}(x) (x+r) (x - r^2) - (\xi_{A_3}(x) + r \xi_{A_2}(x)) x_1^2
\right] \times
$$ 
$$
\left[ \xi_{A_1}(x) (x+r) - {1\over2}(\xi_{A_3}(x) + r \xi_{A_2}(x))
(1-r^2) + {1\over2} (\xi_{A_3}(x) - r \xi_{A_2}(x)) t \right]\times
$$
$$
\Bigl(
(x+r)
\left[\xi_{A_1}(x) (x - r^2) - (\xi_{A_3}(x) + r \xi_{A_2}(x)) (x-r)\right]^2
+
$$
\beq
2 r^2 t \left[\xi_{V_1}(x)^2 (x-r) + \xi_{A_1}(x)^2 (x+r)\right]
\Bigr)^{-1}
\label{Px LT full}
\eeq

\noi
where $C_{H^+}$ is given in (\ref{CH}).  The corresponding expressions
for LR contributions are:

$$ 
P^\perp_l (x) \left(M \stackrel{W_R^+}{\rightarrow} m_{T1}^* l \nu \right) =
C_{W_R^+} {3\pi\over 4}\,
{(x+r) x_1^2 \sqrt{t} \over 2t (x+r) x_1 } \times
$$ 
\beq
{\xi_{A_1}(x) \xi_{V_1}(x) \over
\left[ {3 \over4} \xi_{V_1}(x)^2 (x-r) +
{1 \over4} \xi_{A_1}(x)^2 (x+r) \right]}
\label{Px T1 full}
\eeq

$$ 
P^\perp_l (x) \left(M \stackrel{W_R^+}{\rightarrow} m_{T2}^* l \nu \right) =
- C_{W_R^+} {3\pi\over 4}\,
{(x+r) x_1^2 \sqrt{t} \over 2t (x+r) x_1 } \times
$$ 
\beq
{\xi_{A_1}(x) \xi_{V_1}(x) \over
\left[ {1 \over4} \xi_{V_1}(x)^2 (x-r) +
{3 \over4} \xi_{A_1}(x)^2 (x+r) \right]}
\label{Px T2 full}
\eeq

\noi
where $C_{W_R^+}$ is given by (\ref{C WR}).

To find the average polarization, we must integrate both numerator and
denominator over $x$.  For this, one must know $\xi(x)$.
One possible choice comes from a relativistic oscillator model
\cite{Neubert92},

\beq
\xi(x) = {2r \over (x+r)} e^{- \beta {(x-r) \over (x+r)}}
\label{xi osc}
\eeq

\noi
where $\beta \simeq 1.85$ \cite{Neubert92}.
Another possibility is a monopole approximation,

\beq
\xi(x) = {1 \over 1 + \rho^2 (x-r)/r}
\label{xi monopole}
\eeq

\noi
where $\rho \simeq 1.2\pm 0.25$ \cite{Neubert91}.
For most choices of $\xi(x)$, the integration over $x$ must be done
numerically, but for the monopole approximation with $\rho=1$, one can
obtain reasonably simple analytic expressions (see below).
Since $\xi(x)^2$ appears both in the numerator and denominator of
(\ref{N over N}), \pperpl\ is fairly insensitive to the choice of
$\xi(x)$.
We find that for decays with the lowest value of $r$ ($\sim 0.25$),
the difference between \pperpl\ using  $\xi(x)$ from
(\ref{xi monopole}) for $\rho=1$ (analytic case) and $\rho=1.2$,
and  (\ref{xi osc}) is no more than $15\%$, and considerably less in most
cases (see Fig. 2 and 3).
Thus we use (\ref{xi monopole}) with $\rho=1$ to obtain
the following analytic expressions for the integrals in (\ref{P avg scalar})
and (\ref{P avg LR}):

\bmulteq
I_\perp &=& \int_r^{(1+r^2)/2} dx\, (1-r^2) (x+r) x_1^2 \sqrt{t}\, \xi(x)^2
\nonumber\\
&=& {1 \over 15} r^2 (1-r)(1+6r - 6r^5-r^6)
\nonumber\\
&&\ - \, {2 r (1 -r + r^3 -r^4) \over \sqrt{1 + r^2} }
\tan^{-1}\left( {1-r \over \sqrt{1 + r^2} } \right)\\
&&\nonumber\\
I_S &=& \int_r^{(1+r^2)/2} dx\, (1+r)^2 x_1^3\, \xi(x)^2
\nonumber\\
&=& {1 \over 8} {r^2 (1-r)(1+r)^3 (1+10r^2 + r^4) \over 1 + r^2}
+ {3 \over 2} r^4 (1+r)^2 \ln r\\
&&\nonumber\\
I_L &=& \int_r^{(1+r^2)/2} dx\, (1-r)^2 (x+r)^2 x_1\, \xi(x)^2
\nonumber\\
&=& {1 \over 8} {r^2 (1-r)^3 (1+r) (1+ 8r - 6r^2 + 8r^3 + r^4) \over 1 + r^2}
\nonumber\\
&&\ + \,
2 r^4 (1 -r)^2 \left[\tan^{-1}\left( {2r \over 1 - r^2} \right)
- {\pi \over 2} - {1 \over 4} \ln r \right]\\
&&\nonumber\\
I_T &=& \int_r^{(1+r^2)/2} dx\, 4 t (x+r) x x_1\, \xi(x)^2
\nonumber\\
&=& {1 \over 6} r^2 (1+r)^3 (1+ 3r - 3r^2 - r^3)
\nonumber\\
&&\
+\,
4 r^4 (1 +r^2) \left[\tan^{-1}\left( {2r \over 1 - r^2} \right)
- {\pi \over 2}\right]
+ 2 r^4 (1 -r)^2 \ln r \\
I_{T1} &=& \int_r^{(1+r^2)/2} dx\, 2 t (x+r) x_1 (x- r/2) \, \xi(x)^2
\nonumber\\
&=& {1 \over 12} r^2 (1-r^2) (1+ 3r + 34r^2 + 3r^3 +r^4)
\nonumber\\
&&\
+\,
r^4 (1 +r)^2 \left[\tan^{-1}\left( {2r \over 1 - r^2} \right)
- {\pi \over 2}\right]
+  r^4 (2- r + 2r^2) \ln r\\
I_{T2} &=& \int_r^{(1+r^2)/2} dx\, 2 t (x+r) x_1 (x+ r/2) \, \xi(x)^2
\nonumber\\
&=& {1 \over 12} r^2 (1-r^2) (1+ 9r - 14r^2 + 9r^3 +r^4)
\nonumber\\
&&\
+\,
r^4 \left( 3(1 +r)^2 - 8r \right)
 \left[\tan^{-1}\left( {2r \over 1 - r^2} \right)
- {\pi \over 2}\right]
- 3  r^5 \ln r
\emulteq

\

\



\newpage

\centerline{FIGURE CAPTIONS}

{\bf Fig. 1:} Diagrams which contribute to $M \rightarrow m^{(*)} l \nu$
from (a) the SM $W$ exchange, (b) charged Higgs exchange, (c)
$W_L$-$W_R$ mixing.

\

{\bf Fig 2:} \pperpavgl$/C_{H^+}$ as a function of $r\equiv m/M$ for
$\xi(x)$ given by the monopole approximation (\ref{xi monopole}) with
$\rho =1$ (solid lines), $\rho =1.2$ (dashed lines) and where $\xi(x)^2$
is \naive ly divided out (dash-dot lines).
The top, middle, and bottom sets of curves correspond to
$M \rightarrow m l \nu$,
$M \rightarrow m^*_L l \nu$, and
$M \rightarrow m^* l \nu$ decays, respectively.

\

{\bf Fig. 3:} \pperpavgl$/C_{W_R^+}$ as a function of $r$.
Notation is the same as in Fig. 2,
with the top and bottom sets of
curves corresponding to $M \rightarrow m^*_{T1} l \nu$ and
$M \rightarrow m^*_{T2} l \nu$ decays, respectively.


\begin{thebibliography}{99}
\bibitem{foot semileptonic} In practice,
most models which contribute to \pperpl\ have CPV in both their
quark and lepton sectors,
so that both $d_n$ and $d_e$ are constraints, but it is possible
to construct models which give contributions neither to
$d_n$ nor $d_e$, but which give a sizable \pperpl \cite {Kmu3}.
\bibitem{foot Klmumu}However, one must be careful because \pperpl\ in
models with tree level FCNCs
may be strongly constrained by the $\mu$ longitudinal polarization
in $K_L \rightarrow \mu^+\mu^-$ decays
\cite{Liu Wolf}.
\bibitem{foot SUSY BSG}
If our extended Higgs sector is embedded in a larger theory,
additional particles could appear in the loop and destructively interfere
with the amplitude in (\ref{BSG ampl}),
washing out any potential constraint.
For example, in supersymmetric theories, the charged Higgs
contribution can be cancelled out by loop diagrams containing purely
supersymmetric particles (charginos and squarks), leaving
$B(b \rightarrow s \gamma)$ at or even below the SM prediction
\cite{SUSY BSG}.
\end{thebibliography}
\end{document}